\documentclass[12pt,preprint]{aastex}

\usepackage{graphicx}

\usepackage{natbib}

\newcommand{\be}{\begin{equation}}
\newcommand{\ee}{\end{equation}}
\newcommand{\nn}{\mbox{} \nonumber \\ \mbox{} }
\newcommand{\ba}{\begin{eqnarray}}
\newcommand{\ea}{\end{eqnarray}}

\newcommand{\B}{{\bf B}}

\newcommand{\J}{{\bf J}}
\renewcommand{\v}{{\bf v}}

\newcommand\eg{\textit{e.g.}}
\newcommand\cf{\textit{cf.}}

\newcommand{\Bf}{{magnetic field}}
\newcommand{\Bfs}{{magnetic fields}}

\begin{document}

\author{Maxim Lyutikov\\
Department of Physics, Purdue University, \\
 525 Northwestern Avenue,
West Lafayette, IN
47907-2036 }

\title{Structure of magnetic fields  in non-convective stars}

\begin{abstract}
We develop a theoretical framework to construct axisymmetric  magnetic equilibria in stars, consisting of both poloidal and toroidal magnetic field   components.
In a stationary axisymmetric  configuration,  the poloidal current is a function of the poloidal magnetic flux only, and thus should vanish on field lines extending outside of the star.  Non-zero poloidal current (and  the corresponding non-zero toroidal magnetic field) is limited to  a set   of toroid-shape flux surfaces  fully  enclosed inside the star.  If we demand that there are no current sheets, then on the separatrix delineating  the regions of zero and finite toroidal magnetic field  both the poloidal flux function (related to the toroidal component of the magnetic field)  and its derivative
 (related to the poloidal component)    should  match. Thus, for a given magnetic field  in the bulk of the star,  the elliptical Grad-Shafranov  equation that describes magnetic field  structure inside the toroid is an  ill-posed  problem, with both Dirichlet and Newman boundary conditions and {\it a priori} unknown distribution of toroidal and poloidal electric currents.  We discuss  a procedure which allows to solve this ill-posed problem   by adjusting the unknown current  functions. We illustrate the method by constructing a number of  semi-analytical equilibria connecting to outside dipole and having various poloidal current distribution on  the  flux surfaces closing inside the star. 
In particular, we find a poloidal  current-carrying solution that leaves the shape of the flux function and, correspondingly, the toroidal component of the electric current 
the same as in the case of no poloidal current.  
The equilibria  discussed in this paper may have arbitrary large toroidal magnetic field, and  may include a set of stable equilibria. The method developed here can also be applied to magnetic structure of differentially rotating stars, as well as to calculate velocity field in  incompressible  isolated fluid vortex with a swirl.  
\end{abstract}


\section{Introduction}

Structure of \Bfs\ in non-convective  stars is a long standing issue in astrophysics \citep{Prendergast}.  In fluid non-convective stars,  very general arguments prove that both purely poloidal and purely toroidal \Bfs\ are unstable \citep{Tayler73,Wright,MarkeyTayler,FlowersRuderman}. As a result, \Bfs\ inside  the star should consist of a combination of both   \citep{Prendergast}. An example of such stable (on resistive times scales) equilibrium were found numerically by \cite{BraithwaiteSpruit}.  No  acceptable analytical solution is known.

In this paper we discuss  a semi-analytical  procedure to construct  magnetic structures of fluid stars. We are interested in  developing a  procedure how to construct poloidal-toroidal equilibria of stars, a non-trivial mathematic problem on its own.
We do not discuss the stability of  the resulting configuration. MHD simulations \citep[\eg][]{BraithwaiteNordlund} indicate that stable configurations correspond to stably stratified stars with toroidal magnetic field reaching  inside the star the local value of the poloidal \Bf. 
Thus, we are looking for equilibria where the toroidal and poloidal \Bf\ are, generally, of the same order somewhere  inside the star.

Recent attempts to construct  poloidal-toroidal equilibria of a spheromak-type \citep{BroderickNarayan08,DuezMathis} suffer from a common drawback: they require unphysical surface currents. (Both  works construct spheromak-type  constant-$\alpha$ 
equilibria, which are stable \Bf\ configurations that naturally minimize magnetic energy given a total helicity \cite{Woltjer58}. Such constructions cannot avoid surface current sheets in principle, since poloidal current is non-zero everywhere inside the star, see discussion after Eq. (\ref{jphi})).
In contrast,  our approach ensures that no current sheets are formed inside the star or on the surface.

\section{Magnetic  equilibria: the  Grad-Shafranov equation}

In   MHD equilibria, the  Lorentz  forces are balanced by gradients of pressure and gravitational forces,
\be
\J \times \B =  \nabla p + \rho \nabla \Phi,
\label{MHD}
\ee where $\Phi$ is gravitational potential. 
Dividing Eq. (\ref{MHD}) by $\rho$, taking a  curl, and assuming barotropic fluid, $p=p(\rho)$ (see Appendix \ref{Non-barotropic} for discussion of non-barotropic fluid), Eq. (\ref{MHD}) reduces to 
 \be
\nabla \times  {\J \times \B  \over \rho}= 0
\label{MHD1}
\ee
 
Axially symmetric magnetic  fields can be written as
\be
\B={  \nabla P \times {\bf e}_\phi  + 2 I {\bf e}_\phi \over \varpi }  ,
\label{B0}
\ee
where $2\pi P( \varpi ,z)$ is the poloidal magnetic flux and $ I( \varpi ,z)$ is the poloidal current enclosed
by an axially symmetric loop located at cylindrical coordinates $ \varpi ,z$ \cite[\eg][]{Shafranov}; we set  the speed of light to unity. 
Axial symmetry and the fluid approximation (so the pressure is derivable from a potential and thus cannot have an azimuthal component) give 
$\left. \J \times \B \right|_\phi \propto  ( \nabla I  \times \nabla P ) \cdot {\bf e} _\phi =0$, implying  that poloidal current is a function of $P$ only $I=I(P)$. 
 The force balance equation (\ref{MHD1}) 
 gives
 \be
\nabla P \times \nabla \left(  {\Delta ^\ast P +  { 4 I I'  } \over \varpi^2 \rho} \cdot e_\phi \right) =0,
  \label{GS0}
  \ee
 Taking into account that $I=I(P)$, this reduces to the Grad-Shafranov equation, which in cylindrical coordinates become  \citep[\eg][]{Shafranov}
\be 
 \Delta ^\ast P =  -    \rho  \varpi ^2 F(P)- G(P)
 \label{GS1}
 \ee
 where
\be
\Delta ^\ast  =  \varpi \partial_ \varpi \left( {\partial_ \varpi \over  \varpi }\right)+  {\partial^2\over\partial z^2} 
\ee
is the Grad-Shafranov operator, $ G(P)= 4 I I' $, 
 $I=I(P)$ and $F=F(P)$ are functions of the flux function only. 
 Also note, that toroidal current depends exclusively on the shape of the flux functions:
 \be 
 J_\phi = - {\Delta ^\ast P \over \varpi}
 \label{jphi}
 \ee

As a crucial physical constraint, we will assume that the current cannot leave the star.
One can then identify three regions of space, with different requirements on the functions $I(P)$ and $ \rho F(P)$ (e.~g., \citealt{Monaghan76}, Fig. \ref{BfieldStructure1}).
First, there is the outside vacuum, where $I=0$ and $\rho=0$.
Second, there is a part of the stellar interior  threaded by field lines connecting  to the outside. Since on these field lines  the condition $I(P)=0$ was imposed outside the star, it must still
be true inside; non-zero  density implies that the term containing  $F$ is non-zero inside the star. 
 Finally, if a flux surface is completely closed within the star, on it $I$ and $\rho F$ can both be non-zero. 

Since we are interested in general properties of magnetic equilibria, 
 we make a simplifying assumption of a constant density, $\rho =$ constant. 
 \footnote{Equation of state and initial conditions (\eg\ isentropic or not), are important for the stability of magnetic configurations \citep{BraithwaiteNordlund}. In this paper we are not concerned with the stability, only in finding stationary configurations. The derivation given below can be repeated for any given $\rho(r)$. }
Redefining the function  $F$ to include the assumed constant density, the  Grad-Shafranov equation becomes
\be 
 \Delta ^\ast P =  -  {15 \over 2}  F(P)  \varpi ^2 + G(P)
 \label{GS}
 \ee
 The coefficient $-15/2$ is introduced for numerical convenience, see a comment after Eq. (\ref{FS}). Function $F$ is assumed to be zero outside of the star (since the new definition of $F$ includes density, which is zero outside). Below we refer to function $F$ as the pressure function  \citep[it is related to gradient of presser with respect to flux function][]{Shafranov}  and the function $G$ as the poloidal current function.

\subsection{The mathematical problem}
\label{problem}

Stable equilibria of fluid stars must have considerable toroidal \Bf\ component. As discussed above, the toroidal \Bf\  cannot be non-zero everywhere inside the star, it  must be limited to a  smaller  region of toroidal shape (in case of axial symmetry). On the separatrix between the regions of non-zero toroidal \Bf\  and  the bulk (which has  zero  toroidal \Bf),  both the poloidal and the toroidal components  of  the \Bf\ must connect smoothly, without a surface current. For axisymmetric configurations, the toroidal components of  the \Bf\  is related to flux function, while the poloidal component of the \Bf\ is related to the gradient of the flux function, see Eq. (\ref{B0}) with $I=I(P)$.  In order to avoid  current sheets, on the separatrix  both the flux function $P$ and its  normal derivative $\partial_n P$ should be continuous. 

Thus, we are faced with an unusual mathematical problem: we need {\it to find solutions of an elliptical equation (\ref{GS}) 
while   matching on a given boundary both the flux function (equivalently, the toroidal B-field) and the  flux function normal derivative (equivalently, the poloidal B-field).  }
This makes an elliptical equation (\ref{GS}) is formally  over-constrained, as it should satisfy both Newman and Dirichlet boundary conditions simultaneously. In addition,
 the functions $F$ and $G$ are unknown{ \it a priori} and should be found as a part of the solution. Thus, we are trying to solve an  over-constrained equation with two unknown functions, which themselves are functions of the solution only. 

For a given \Bf\ structure in the bulk (with no 
poloidal current), we are looking for  \Bf\ structure inside a  poloidal current-carrying toroid of a given shape.
In principle, the shape of this toroid also is not known {\it a priori}. But this does not create any addition mathematical complications: for any shape of the boundary, the solutions inside and outside of the toroid should be matched satisfying both Newman and Dirichlet boundary conditions simultaneously. (As a side note, a one-dimensional problem with both Newman and Dirichlet boundary conditions can be solved if there exist several eigenvalues  for  the solution of  either Newman or Dirichlet problem. The four corresponding boundary conditions can be satisfied by choosing two integration constants and two eigenvalues. This method does not work for the  two dimensional domain, since the shapes of the flux functions corresponding to different eigenvalue problems are different.)
In this paper we devise a procedure that allows to built  both  the functions  $F$ and $G$  and the flux function $P$ inside the toroid  for given boundary conditions  at the boundary of the toroid.

\section{Matching the  bulk and the  toroid solutions}

\subsection{Solution in the bulk}

\begin{figure}[h!]
\includegraphics[width=\linewidth]{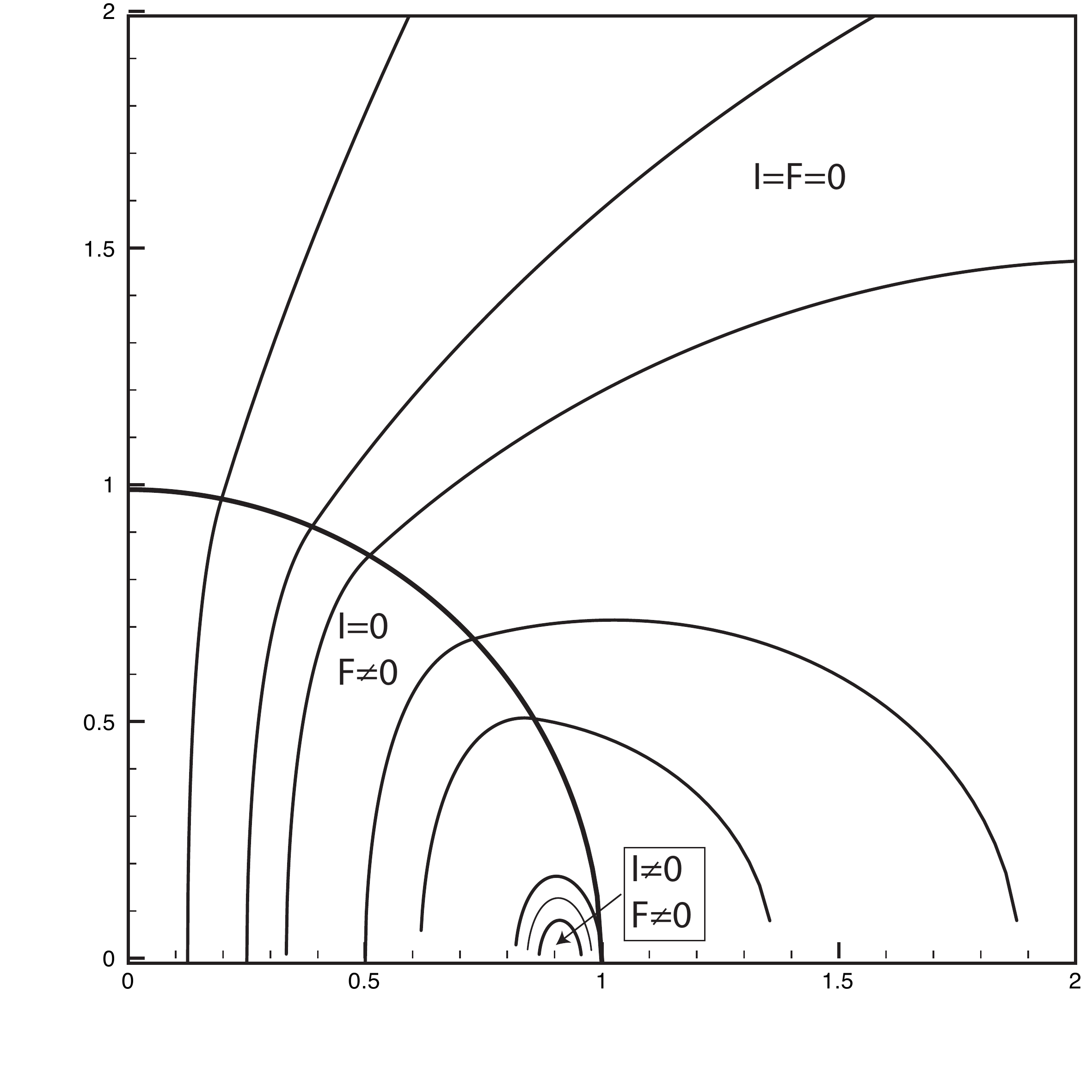}
\caption{Structure of \Bf\  for constant density. Outside solution correspond to vacuum dipole. Inside, there is a set of nested flux surfaces near the surface and near the equatorial plane. Closed flux surfaces must  carry poloidal current for the structure to be stable.}
\label{BfieldStructure1}
\end{figure}

As we discussed in \S \ref{problem}, we are faced with an unusual  mathematic problem of continuously  matching  the solution of the elliptical  Grad-Shafranov equation (\ref{GS}) with zero poloidal current   ($I=0$) and some distribution of the toroidal current ($F\neq 0$) in the bulk and unknown non-zero poloidal current ($I\neq 0$) and some toroidal current inside a toroid.
As a basic solution {\it in the bulk} we take  a well known dipolar, purely poloidal solution connecting to outside dipole \citep{Ferraro54,Shafranov} (the procedure developed below can be repeated  for any shape of the enclosed toroid)
\ba &&
P_0 = { ( 5 R^2 - 3 (\varpi^2 - z^2)) \varpi^2 \over 4 R^2} B_0
\nn &&
\B = {B_0\over 2R^2}[3  \varpi z  {\bf e}_\varpi + ( 5 R^2 - 6 \varpi^2-3 z^2){\bf e}_z]
 \label{FS}
\ea
This solution corresponds to $I=0$ and  $F=1 $. 
It has   a set of flux surfaces closing inside a star, see Fig. (\ref{BfieldStructure1}). 
Under the fluid approximation, this solution is unstable to non-axysimmetric perturbations \citep{Tayler73,FlowersRuderman}.

The flux surfaces are given by solution  $z(\varpi)$ of (\ref{FS}) for $ 0< P < (25/48)$.  (Below, for conciseness we set $R=1$, $B_0=1$.) For a given meridional plane  they satisfy
\be
{z^2 + \varpi^2 } = {5 \over 3} - {4\over 3} {P \over  \varpi^2}
\ee
There is a set of nested  toroids enclosing  within a star and centered on the apex point
$ \varpi= \sqrt{5/6} $ and $z=0$.
The ellipticity of the boundary is $1/2$ (the flux surfaces are not ellipses, though). 
The first closed flux surface  (saparatrix) corresponds to $P_s =1/2$ and  equation for it can be written as
\be
\left ( {z } \right)^2 = {5\over 3}  - { 2 \over 3 \left (  \tilde{x} + 2 /\sqrt{5}  \right)^2 } - \left(  \tilde{x} + 2 /\sqrt{5}  \right)^2
\label{closed1}
\ee
where  $ \tilde{x} =  \varpi /R - 2/\sqrt{5}$ are coordinates centered on the center line of the enclosed
toroids, corresponding to $\varpi= (2/3)^{1/4}$, $z=0$ (since the flux surfaces are not ellipses, the symmetry axis of the separatrix does not intersect the apex). The separatrix intersects the  equatorial plane at points $\varpi=\sqrt{2/3}$ and $\varpi= 1$.
The maximum value of $P$   equals $P_{max} = (25/48)$. This value is {\it higher} than the maximum value  for vacuum dipolar field
at a given radius $P_s =1/2$; this   allows a smooth sewing of the   vacuum dipole potential, which is a decreasing function of radius, with increasing  potential at small radii of the internal solution.

\subsection{The  algorithm to calculate \Bf\ structure inside the toroid}

The solution (\ref{FS}) has a set of flux surfaces closing inside a star. Toroidal \Bf\ should be confined to this region. On the separatrix between the regions of zero and finite  \Bf, both the flux function $P$ and its derivative should be continuous. In this section we discuss how the unknown functions $F$ and $G$ can be adjusted to satisfy the over-constrained boundary conditions on the separatrix $\partial$.

We  expand the unknown functions  $F(P)$ and $G(P)$ in terms of their argument $P$ near the separatrix $\partial$, corresponding to  $P=P_0$ (subscript $0$ corresponds to quantities evaluated on the  separatrix)
\be \left(
\begin{array}{l}
F\\
G
\end{array} 
\right)= 
\sum_{j=0}  {(P-\left.P_0\right|_\partial)^{j} \over j!} 
 \left(
\begin{array}{l}
F^{(j)}(\left.P_0\right|_\partial)\\
G^{(j)}(\left.P_0\right|_\partial)
\end{array}
\right)
\label{FG}
\ee
Here $F^{(j)}(\left.P_0\right|_\partial)$ and $
G^{(j)}(\left.P_0\right|_\partial)$ are values of the derivatives on the separatrix. These are numerical coefficients to be determined. The upper bound in the sum is determined by the number of initial starting points and the imposed  smoothness order of the solutions at the points of intersection. The first terms in the series (\ref{FG}) are known:
at the separatrix $F=1$, $G=0$.

At each point inside  the separatrix the function $P$ can be expanded in Taylor series  in a following manner.
On the separatrix the flux function  is constant $P=\left.P_0\right|_\partial$,  while its derivative is normal to the separatrix and is a known function, $\partial_n \left.P_0\right|_\partial$. This allows us to express from Eq. (\ref{GS}) the  second normal derivative at the boundary $\partial$ as some function $ {\cal F}$ of $P_0, \,\partial_n P_0, \,F(\left.P_0\right|_\partial)$ and $G(\left.P_0\right|_\partial)$:
\be
\left. \partial_n^2 P \right|_\partial
 = {\cal F}\left(\left.P_0\right|_\partial, \partial_n \left.P_0\right|_\partial, F(\left.P_0\right|_\partial),G(\left.P_0\right|_\partial)\right)
\label{1}
\ee
 Here $F(\left.P_0\right|_\partial)$ and $G(\left.P_0\right|_\partial)$ are unknown coefficients to be fitted.
By taking derivatives of Eq. (\ref{GS}) along the normal to the separatrix we can express higher order derivatives $\partial_n^{2+m} P $ as functions
of the given $\left.P_0\right|_\partial$ and $ \partial_n \left.P_0\right|_\partial$, as well as expansion coefficients  $F^{(m)}(\left.P_0\right|_\partial)$ and $G^{(m)}(\left.P_0\right|_\partial)$. 
The solution inside the toroid  can then be represented in Taylor expansion in terms of derivatives at the boundary:
\be
P({\bf r})= \sum_j {s ^j  \over j!} \left.  \partial_n^{(j)} P_0\right|_\partial
\ee
where $s$ is a distance along the normal to a given point ${\bf r}$ inside the separatrix.  The values of $\left.  \partial_n^{(j)} P\right|_\partial$ can then be expressed in terms of unknown coefficients $\partial_n^j\left. F \right|_\partial $ and $\partial_n^j\left. G \right|_\partial $, similarly to  Eq. (\ref{1}) for $\left. \partial_n^2 P \right|_\partial$. An  explicit example of the implementation of this  procedure is given in Appendix \ref{expansion}. 

Next, we  impose a condition that values of  $P$ extrapolated from different  points on the boundary converge smoothly on a chosen set of  points inside the toroid. Though
 generally this convergence is not guaranteed, we can consider  extrapolation from  a finite number of points  on the boundary and limit the expansion to  the appropriate  number of derivatives  $F^m(\left.P_0\right|_\partial)$ and $G^m(\left.P_0\right|_\partial)$ to ensure smoothness of $P$ at the convergence points to a given order.  We will require smoothness of $P$ up to second order; this ensures the absence of current sheets. Thus,  matching of the expansions of the flux functions from different points  gives the coefficients  $F^{(m)}(1/2)$ and
$G^{(m)}(1/2)$. This  determines simultaneously 
both the  flux function $P$ and functions $F$ and $G$. Since this procedure can be extended to the arbitrary expansion order, it determines the functions $P$,   $F$ and  $G$ to arbitrary precision.

The procedure described above is implemented starting from three particular points on the toroid corresponding to the symmetry axes of the separatrix: at the points where the separatrix  intersects the magnetic equator $z=0$,  ${\bf r}= \sqrt{2/3} R {\bf e}_\varpi$ and ${\bf r}=R {\bf e}_\varpi$, as well as at the point where \Bf\ on the separatrix   is parallel to the equator,   ${\bf r} =(2/3)^{1/4} R  {\bf e}_\varpi + \sqrt{ (5- 2 \sqrt{6})/3} {\bf e}_z$.
Thus, the point where the integration curves intersect is $\varpi= (2/3)^{1/4}$ and $z=0$. 
Note, that the point where \Bf\ on the separatrix  is parallel to the equator has different cylindrical radius than the apex point of the enclosed cylinders; enclosed 
flux surfaces are not ellipses. This is true both for the solution (\ref{FS}) as well as the numerical solutions found here.

Since the procedure described above is not unique, we expect  various internal structures of the enclosed toroid. We have two unknown function, $F$ and $G$: choosing one of them determines the other. This will create a set of various equilibria.   Below we describe two procedures: first expansion of functions $F$ and $G$ in terms of $P$ up to a given order, and secondly specifying $F$ and solving for $G$. As is shown in \S \ref{3}, simultaneous  expansion of functions $F$ and $G$ in terms of $P$ (as opposed to arbitrary specifying one of them) selects a particular class of solutions, when the toroidal current and the form of the flux functions remain the same as in the case of zero poloidal current. (Recall that  the  toroidal current depends only on the shape of the flux function, Eq. (\ref{jphi}).)

\subsection{ Expansion of $F(P)$ and $G(P)$ in terms of $P-\left.P_0\right|_\partial$}
\label{3}

For three starting points on the separatrix, there are five matching  conditions: equality of flux functions (two conditions),  equalities of first and second  radial derivative, and zero derivative along the direction normal to the equatorial plane. We expand the functions $F$ and $G$ up to the third order in $P-\left.P_0\right|_\partial$:  boundary conditions require $G=0$ and $F=1$ on the boundary; then five expansion parameters are determined from the matching  conditions leaving one free parameter. The free parameter normalizes the overall strength of the poloidal current inside the toroid.

\begin{figure}[h!]
\includegraphics[width=.47\linewidth]{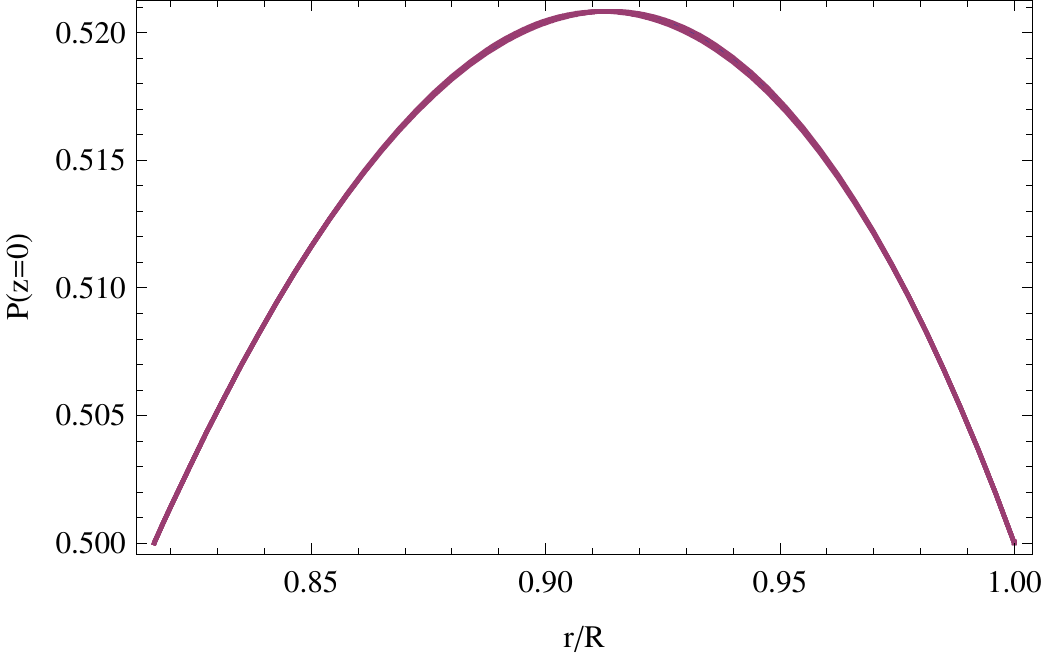}
\includegraphics[width=.49\linewidth]{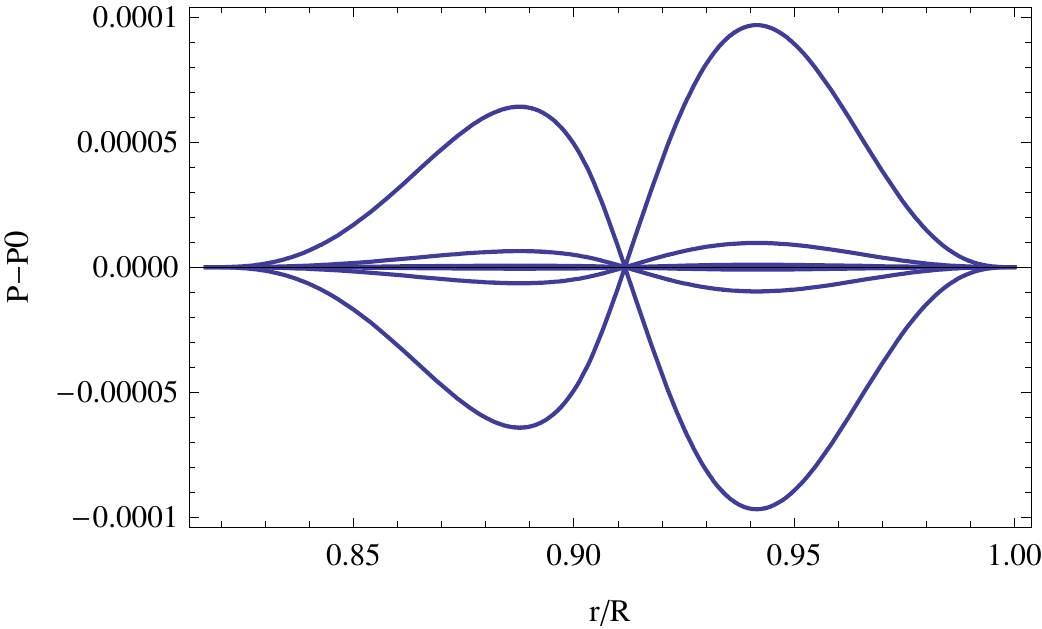}\\
\includegraphics[width=.47\linewidth]{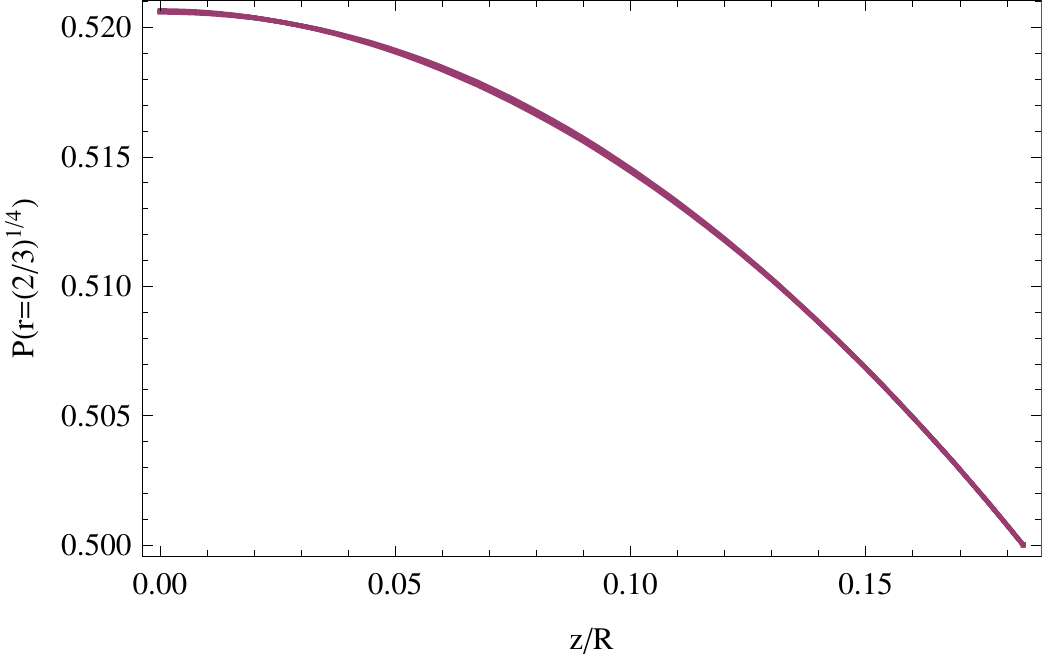}
\includegraphics[width=.49\linewidth]{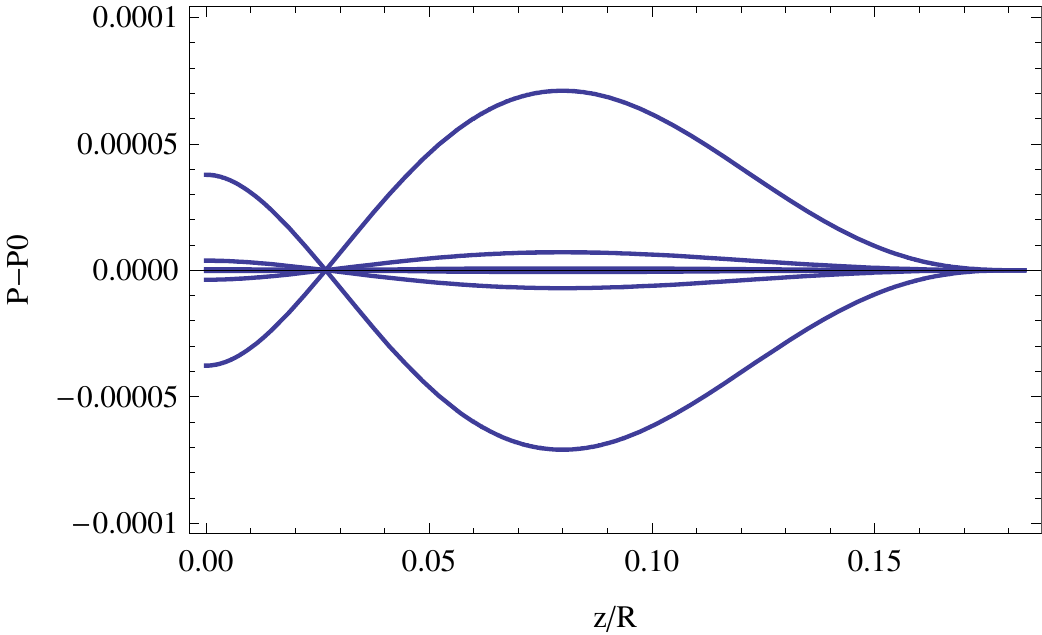}
\caption{Structure of enclosed flux function $P$ for the case when functions $G$ and $F$ are expanded in terms of the flux function at the separatrix. 
{\it Upper row}: Value of the flux function in the equatorial plane within the submerged toroid (upper left panel). At this scale different curves overlap.
Upper right panel:  difference between the poloidal current-carrying solutions and poloidal current free solution (\ref{FS}) in the equatorial plane. 
{\it Lower row}:  Value of the flux function along the vertical symmetry axis of the submerged toroid, $\varpi = (2/3)^{1/4}R$  (low left panel). Lower right panel:  difference between the poloidal current-carrying solutions and poloidal current free solution (\ref{FS}) along $\varpi = (2/3)^{1/4}R$.}
\label{main1}
\end{figure}

\begin{figure}[h!]
\includegraphics[width=.32\linewidth]{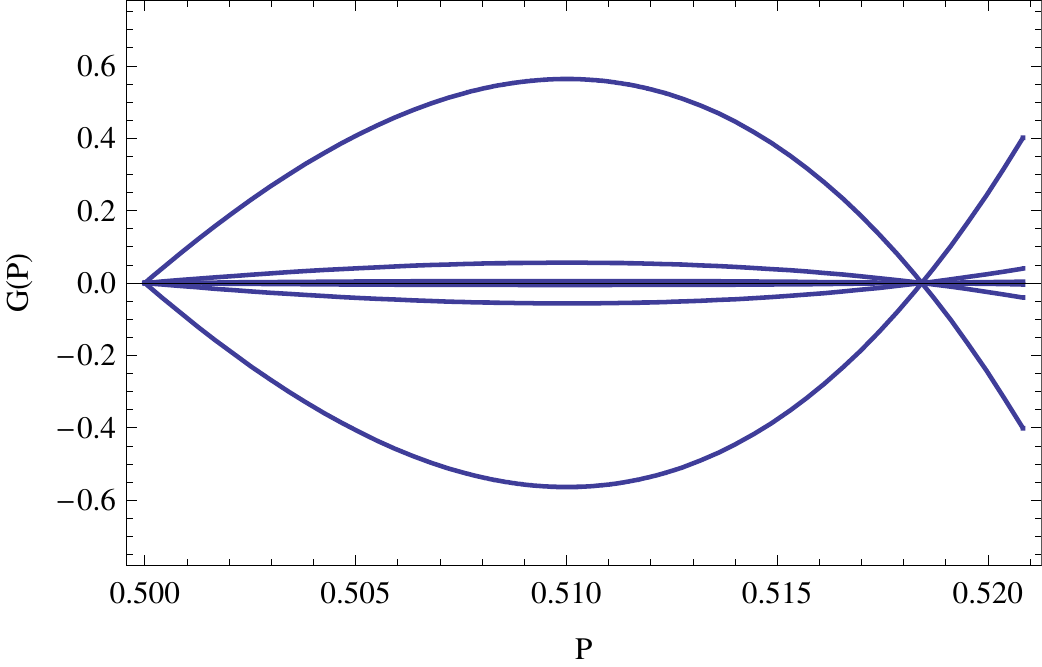}
\includegraphics[width=.32\linewidth]{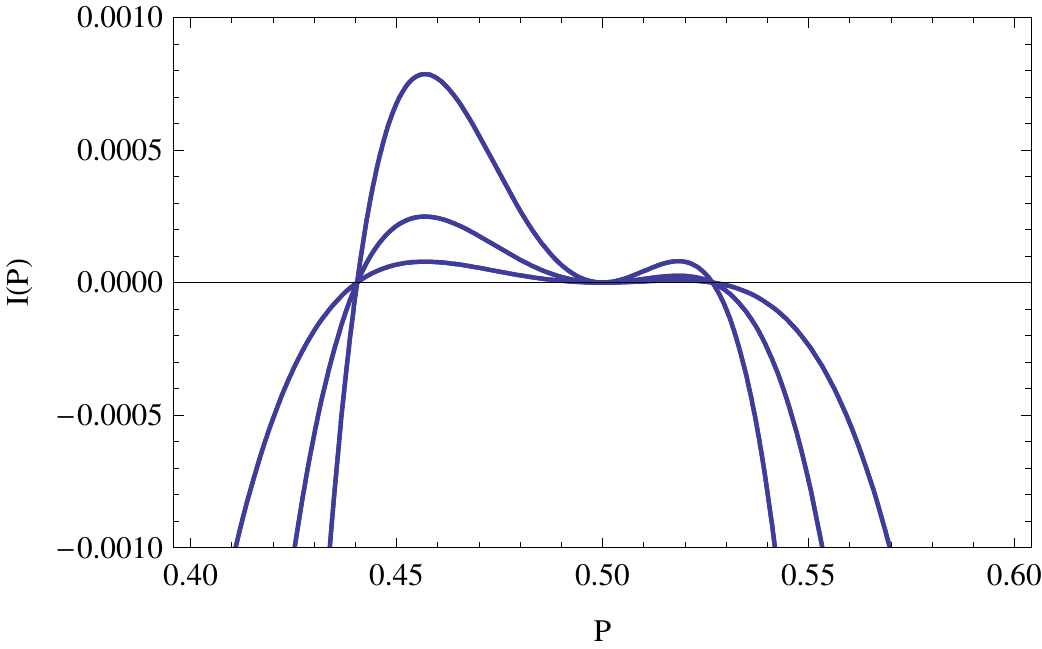}
\includegraphics[width=.32\linewidth]{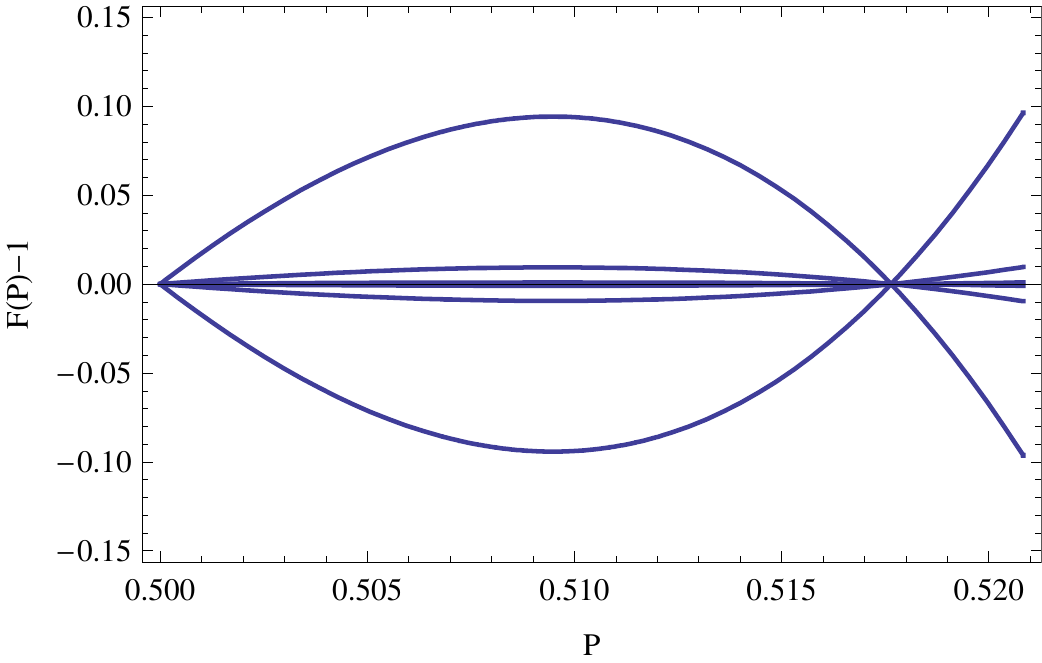}
\caption{Structure of current and pressure functions  for the case when functions $G$ and $F$ are expanded in terms of the flux function at the separatrix.  {\it  Left panel}: Function $G=4 I I'$;  {\it Center panel}: current function $I(P)$;  {\it Right panel}:   function $F(P)-1$. This figure illustrates the value of the right hand side terms in Eq. (\ref{GS}) for different values of the amplitude of the poloidal current. }
\label{main}
\end{figure}

Results of calculations are presented in Fig. \ref{main1}-\ref{main}. Somewhat surprisingly, in case of non-zero poloidal current $G\neq 0$ and non-constant pressure function $F\neq 1$, the algorithm leaves the shape of the flux functions the same as for zero poloidal current, $P(r,z) \approx P_0(r,z)$, see Eq. (\ref{FS}) and Fig. \ref{main1}. At the same time, the  functions $F$ and $G$ on the right hand side of Eq. (\ref{GS}) are adjusted  to match the zero poloidal current case,
$-(15/2) F(P)  \varpi ^2 + G(P) \sim -(15/2)   \varpi ^2$,  Fig. \ref{main}.  

Thus, the toroidal current and  the shape of the poloidal \Bf\ remains the same. In Fig.  \ref{main}  the current-carrying flux functions closely resemble  the current-free one, with precision of $\sim 10^{-4}$. We remind that this  calculation is based on only three extrapolation points. In comparison,  when the pressure function  $F$ is imposed, \S  \ref{F}, the distortion of the form of the flux function is of the order of unity.  

\subsection{Other solutions}
\label{F}

Next  we  find the flux function $P$ and   coefficients  $G^m(P_0)$
assuming that inside the toroid $F$ is a given function of $P$.

\subsubsection{Constant  $F\neq 1$    }

Let us  find the flux function $P$ and   coefficients  $G^m(P_0)$
assuming that inside the toroid $F=\xi \neq 1$.  We expand solutions to 5th order derivative near the separatrix in order to match expansion at the equatorial plane to the second order (three coefficients), match it to expansion along $z$ axis, and condition of vanishing radial \Bf\ at the equator. We investigate solutions as functions of the parameter $\xi$.

\begin{figure}[h!]
\includegraphics[width=.49\linewidth]{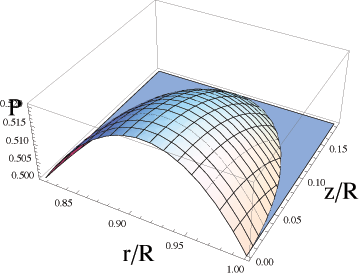}
\includegraphics[width=.49\linewidth]{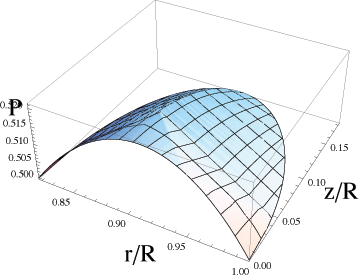}  \\
\includegraphics[width=.49\linewidth]{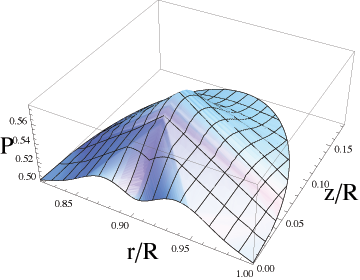}
\includegraphics[width=.49\linewidth]{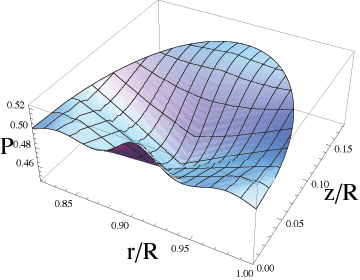}
\caption{Shapes of the flux function $P$ for given $F=\xi 1$ . {\it Upper Left Panel}: analytical solution Eq. (\ref{FS}). {\it Upper Right Panel}: numerical  solution for $\xi=1$. In this case the numerical solution closely matches the analytical solution Eq. (\ref{FS}); this vindicates the fitting procedure. {\it Lower Left Panel}:  $\xi=10$. {\it Lower Right Panel}:  $\xi=- 10$ }
\label{Fa}
\end{figure}

\begin{figure}[h!]
\includegraphics[width=.49\linewidth]{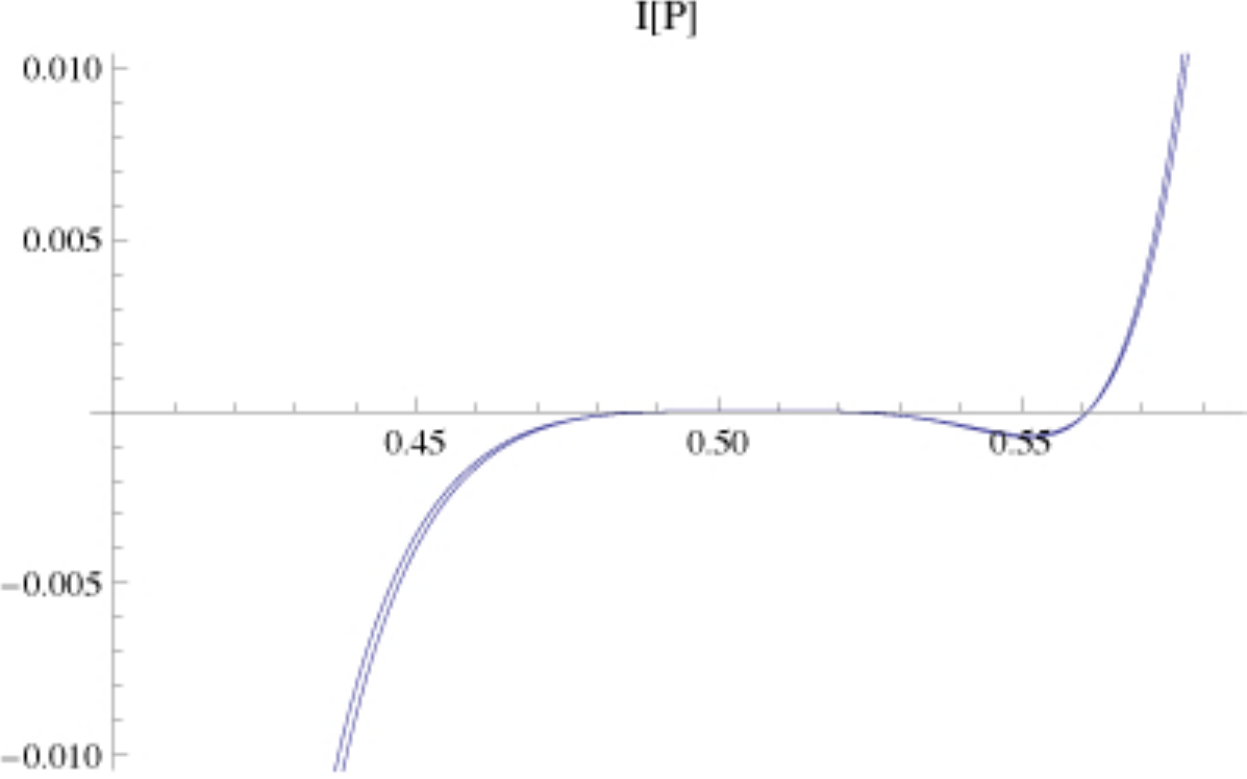}
\includegraphics[width=.49\linewidth]{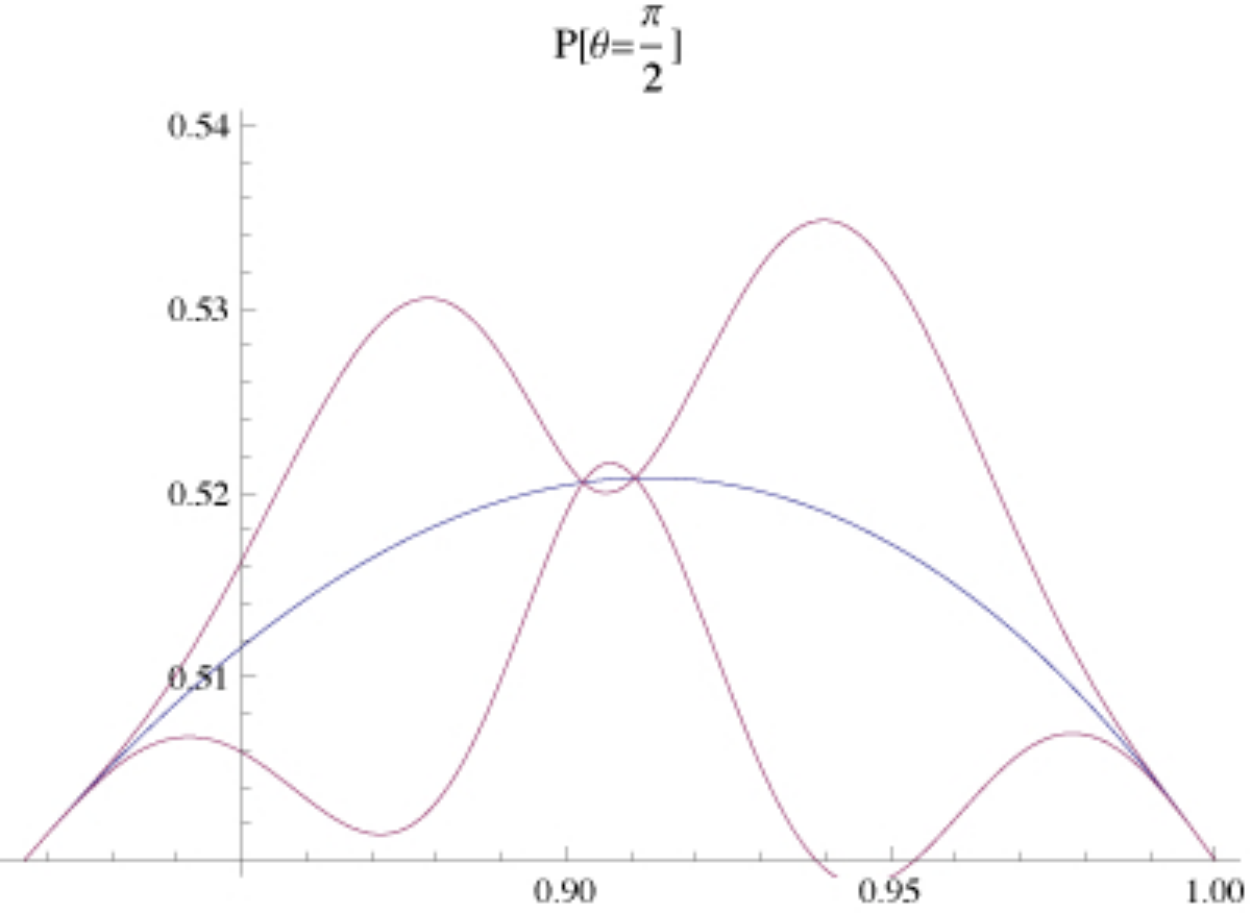}  \\
\includegraphics[width=.49\linewidth]{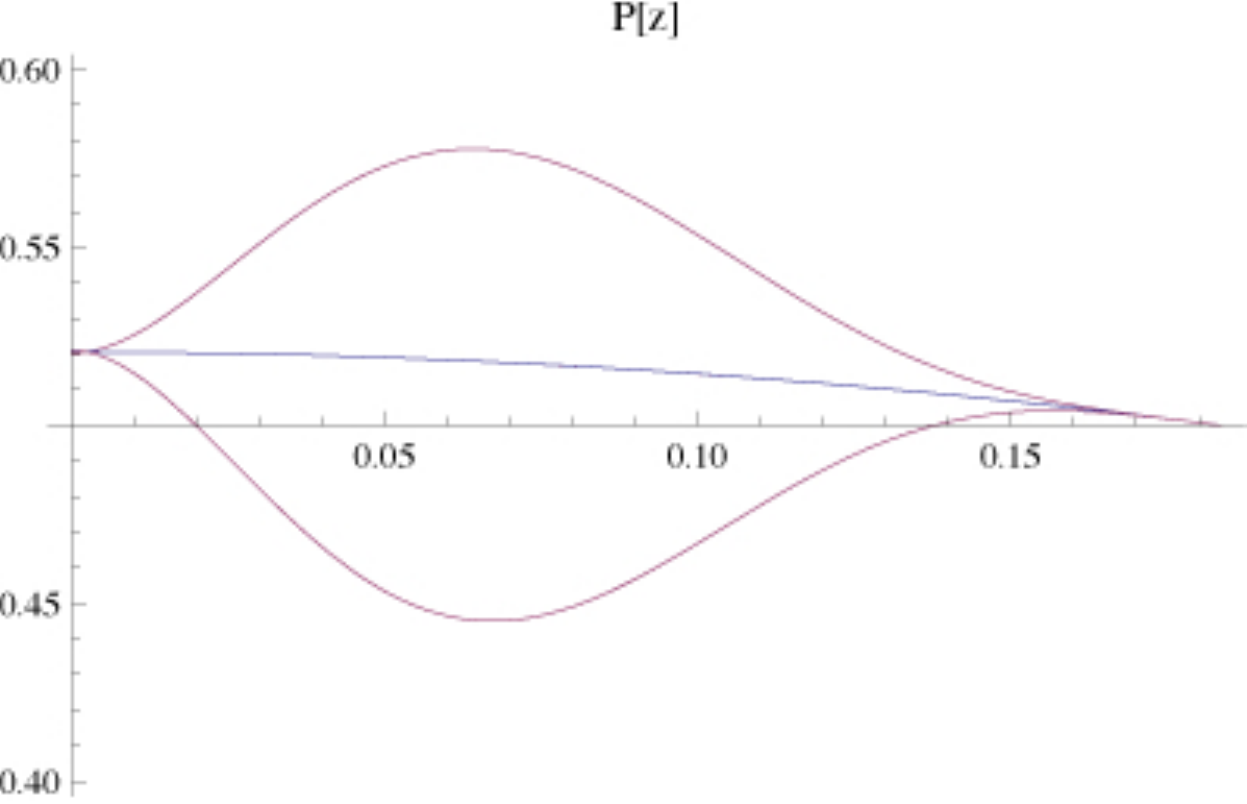}
\includegraphics[width=.49\linewidth]{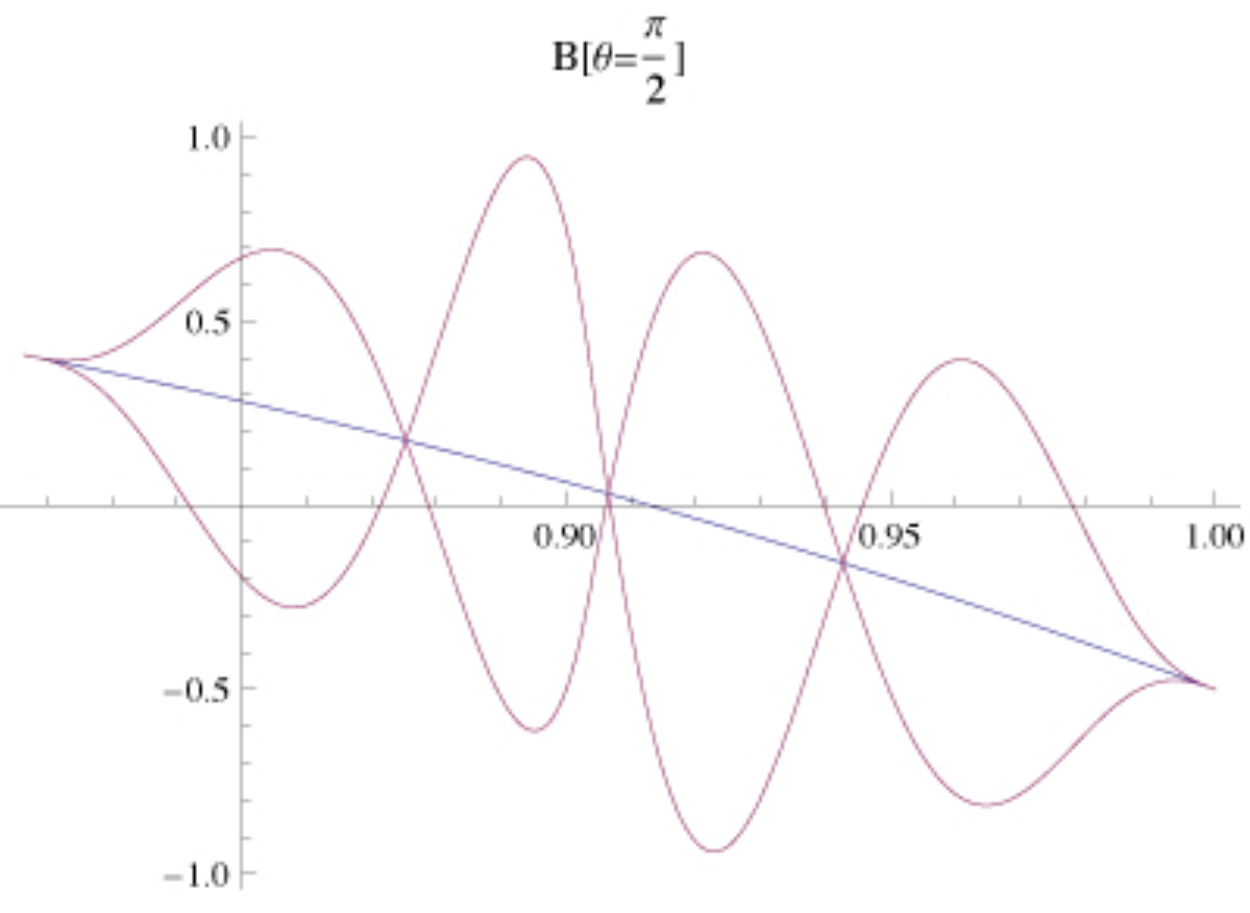}
\caption{Current as function of flux function $P$, $ P$ in the equatorial plane, $P$ along the line $\varpi =(2/3)^{1/4}$ as function of $z$ , $B_z$ in the equatorial plane for different values of parameter $\xi=1, \pm 10$ (except in the $I(P)$ plot where  $\xi=1, \pm5, \pm  10$.  The middle line in the plots (b-d) is both the analytical solution (\ref{FS}) and numerical solution for $\xi=1$ (the difference between them is tiny on this scale).}
\label{IofP}
\end{figure}

First,  for $\xi =1$ the procedure described above closely reproduces the solution (\ref{FS}), see Figs. \ref{Fa} and \ref{IofP}.  
Thus,  if we keep  the function $F$  the same in the bulk and inside the toroid, the procedure described above converges to the solution $G=0$.

\begin{figure}[h!]
\vskip -1 truein
\includegraphics[width=\linewidth]{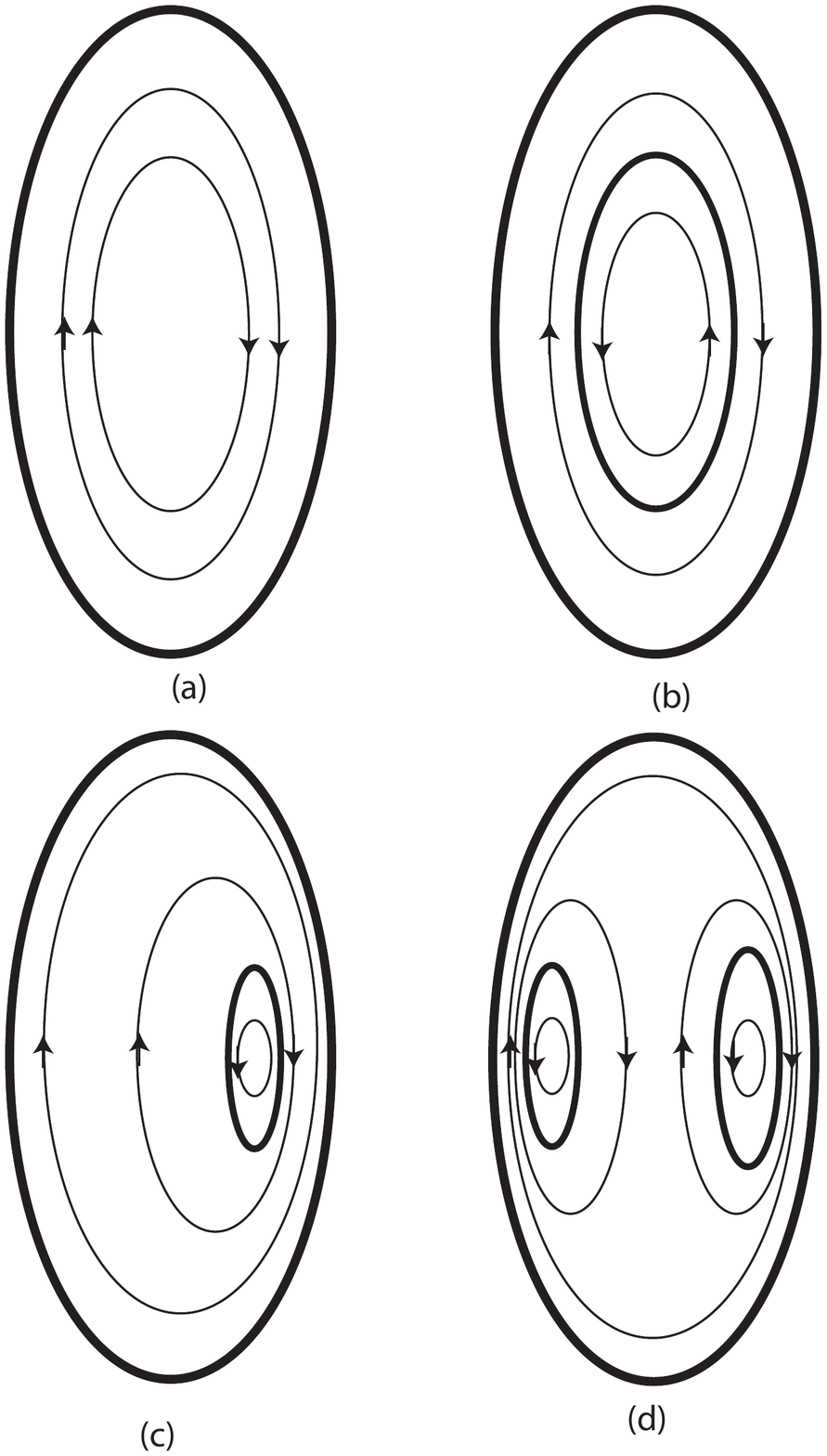}
\caption{Qualitative structure of the poloidal flux surfaces  closing inside a star for $F=\xi\neq 1$. For $ -3.4 <\xi < 1.99$ ({\it Panel (a)}) the direction of the poloidal \Bf\ inside the toroid  is of one sign. For very large,  $\xi \gg 1$ ({\it Panel (b)}) there is a reversal of \Bf\ close to the axis of the toroid. For large negative,  $\xi \ll  -4.1 $ ({\it Panels (c,d)}) the reversal of \Bf\ occurs inside two separatricies not aligned with the main axis.  Thick lines indicate separatricies, where \Bf\ reverse direction. These figures are illustrations only.}
\label{toroids}
\end{figure}

For $\xi \neq 1$, there is non-zero poloidal current and, correspondingly  non-zero toroidal \Bf, Figs. \ref{Fa}.
For large positive or large negative $\xi$, the topology of the \Bf\ inside the toroid changes,  see Figs.   \ref{Fa}, and \ref{IofP} and \ref{toroids}.
For $\xi > 1.99$ there forms a new set of flux surfaces centered close to the axis of the toroid  where the poloidal field reverse.
For $-4.1 \leq \xi \leq -3.4$ there is an off-centered set of flux surfaces where the poloidal field reverse. 
For $ \xi \leq - 4.1$ additional  off-centered set of flux surfaces forms. For large absolute values of $|\xi|$ the sizes of the internal toroids increase, while no other change of topology occur.  

The plot of $I(P)$ indicates that when solutions extend to values  of the flux function large than, approximately $.6 $ (in dimensionless units $R=B_0=1$) or smaller that $0.4$ for the different sign of $F$, the toroidal field, $\propto 2 I /r$ approaches poloidal field ($=1/2$ at equator). (In case of $I=0$ the maximum value of  $P$ is $25/48$.) The maximum value of $P\sim 0.6$ or minimum  $P\sim 0.4$ is reached (approximately) for $|\xi| \geq 10$. Thus, we expect that for  $|\xi| \geq 10$ the resulting configuration is stable.

Some of the found configurations have poloidal \Bf\ reversing inside the  submerged toroid. Though such configurations have not so far been seen in simulations, we hypothesize that this may be related to initial conditions used in simulations.

\subsubsection{ $F= 1+\alpha (P-1/2) $    }

\begin{figure}[h!]
\includegraphics[width=.32\linewidth]{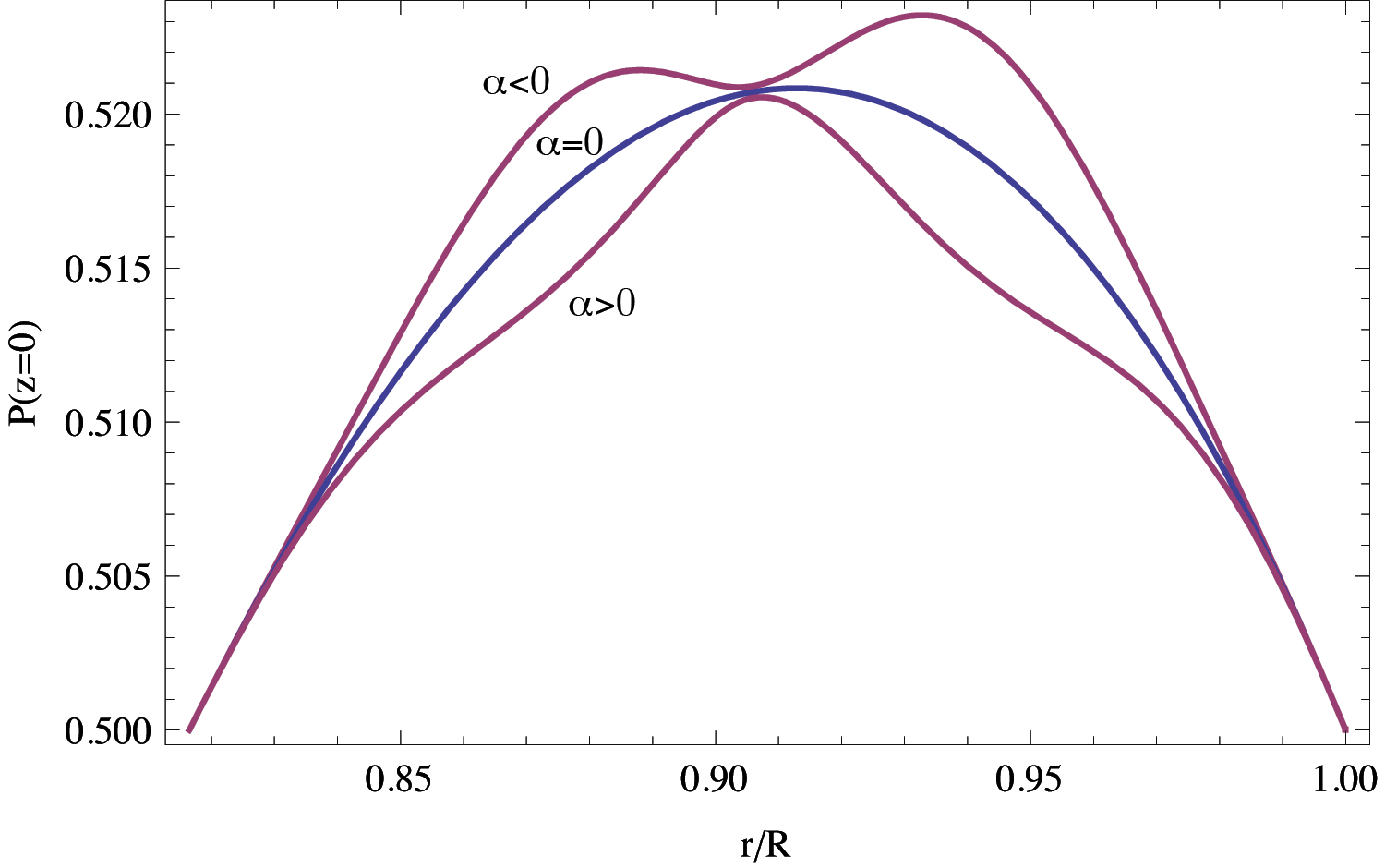}
\includegraphics[width=.32\linewidth]{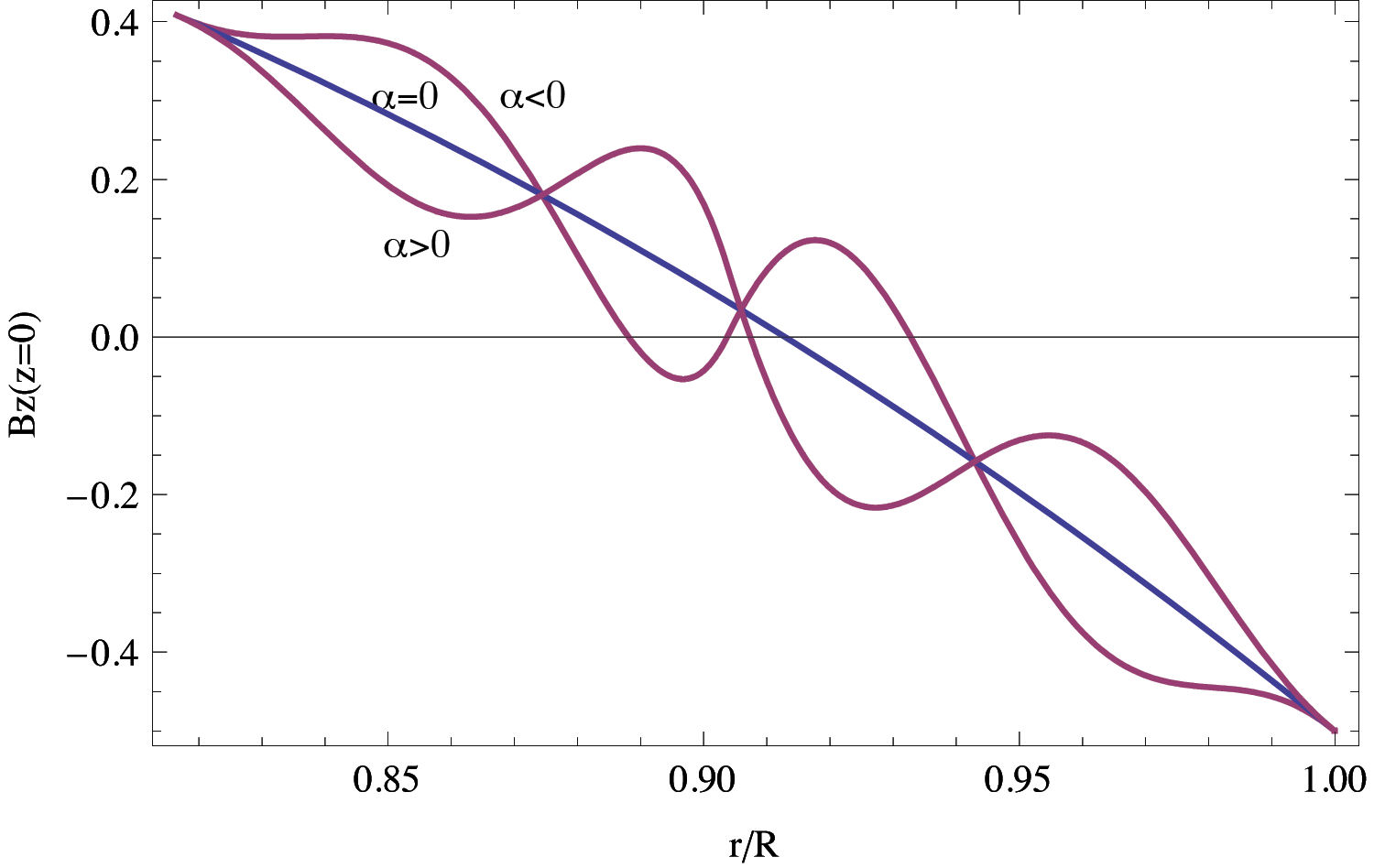}
\includegraphics[width=.32\linewidth]{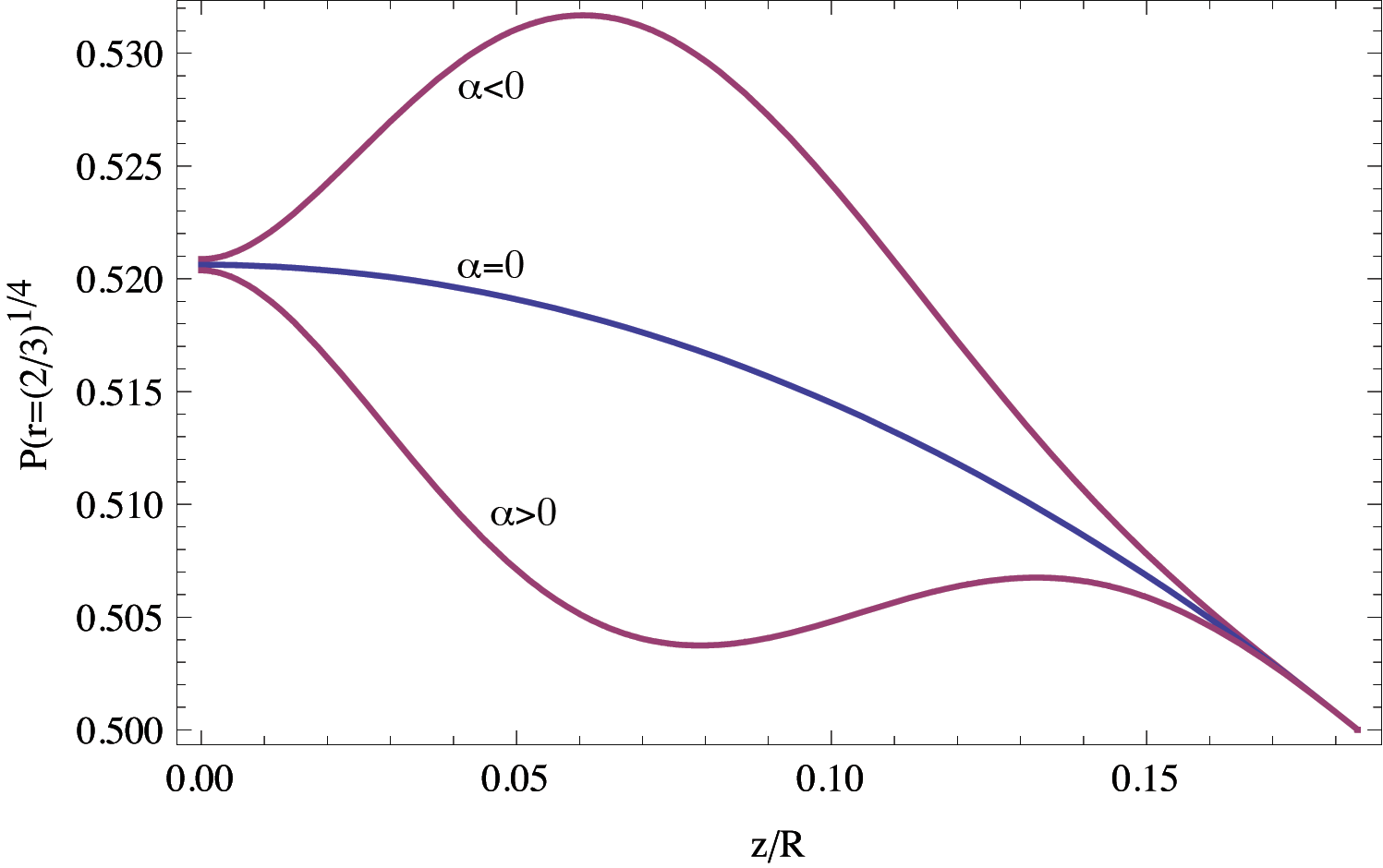}
\caption{Form of the flux functions and \Bf\  for a given  $F=1+\alpha (P-1/2)$. Qualitatively,  solutions remain the same as in the case $F=\xi \neq 1$.
{\it  Left panel}: values of $P$ at the equatorial plane. {\it Center panel}: values of \Bf\ at the equatorial plane. {\it Right panel}:  values of $P$ at $\varpi = (2/3)^{1/4}$.
Qualitatively, the structure of the solutions remain the same as in Figs. \ref{IofP}-\ref{toroids}}
\label{alpha}
\end{figure}

Solutions corresponding to $F= \xi \neq 1$ has a disadvantage that the toroidal current density experiences a jump on the separatrix. If we impose a smooth variation of toroidal current at the separatrix, \eg\ in the form $F= 1+\alpha (P-1/2) $, the behavior of the solutions qualitatively remains the same, Fig. \ref{alpha}.

\section{Conclusion}

In this paper we develop a procedure to construct poloidal-toroidal equilibria of fluid non-convective stars. This involves a mathematically unusual procedure to  find the distribution of unknown poloidal and toroidal currents, which depend only on the shape of the magnetic flux function,  and the shape of the magnetic flux function itself; all these quantities  must  match the  shape of the separatrix and specified Dirichlet and Newman boundary conditions on it. The procedure we propose uses expansion of the current functions in powers of the flux function to a specified  order, which depends on the number of points used to extrapolate the flux function. Generally, this procedure is not unique; for example, given some pressure function $F$ it will determine the poloidal current function $G$, or vise versa.  This is qualitatively consistent with results of numerical simulations \cite{BraithwaiteNordlund} which  show that depending on the initial conditions a variety of final states are achieved.

One particular configuration stands out among the all possibilities. It corresponds to simultaneous expansion of two functions $F$ and $G$ in terms of the value of the flux function $P$ on the separatrix. 
In this particular case,  the shape of the poloidal current-carrying flux surfaces remains the same as in the case of no poloidal current: newly found functions $F$ and $G$ add up to produce the toroidal current of the poloidal current-free configuration.

An important feature of our solutions is that there no current sheets, neither   on the surface of the enclosed toroid nor on the surface of the star. On the other hand,  
we do not test for the stability of the resulting magnetic structures. This could be done, in principle, by minimizing magnetic energy at a given helicity for \Bf\ structure inside the toroid. Results of numerical simulations  \citep{BraithwaiteNordlund} indicate that stable \Bf\ configurations  require both stable stratification and typically have toroidal \Bf\  of the order of the poloidal \Bf. Our procedure, in principle, allows arbitrary ratios of toroidal and poloidal \Bfs, and we expect that some of the found \Bf\ configurations to be stable.

One of the main limitations of our approach is that we assume barotropic equation of state. In contrast,  in stably stratified stellar interiors buoyancy forces play an important role, dominating typically over magnetic forces. Still, even in non-barotropic fluid the poloidal current is still function of $ P $ (Appendix \ref{Non-barotropic}), so that  the toroidal \Bf\  is still limited to a set of fully submerged  flux surfaces. On the separatrix of the regions with vanishing and finite toroidal \Bf, again, both the flux function and its derivative should match. We leave this problem to future considerations. 

The method developed here can also be applied to rotating stars  when  the magnetic 
axis and the rotation axis are aligned. In this case the Grad-Shafranov equation has a form similar to (\ref{GS}) with  $F \rightarrow 
F+  \varpi^2 \Omega \Omega'$, where $\Omega(P)$ is the angular velocity of rotation, which is constant on a given flux surface 
\citep[][]{Chandrasekhar1956ApJ...124..232C}. The new function $\Omega(P)$  can be expanded near the saparatrix in a same way as the pressure $F$ and the current function $G$.

There is  a  hydrodynamical analogue to the  solutions presented here, describing an isolated fluid vortex with a swirl.  
Stationary magnetic configurations are related to  velocity field of a time-independent flow of  an incompressible fluid, with the substitution $\v \rightarrow \B$ \citep{Shafranov}. In case of a steady, axially symmetric fluid motion, the velocity potential satisfies the same equation as (\ref{GS1}), the so-called Bragg-Hawthorne and/or Squire-Long equation  \citep[see, \eg][Eq. 165.13]{Lamb}. Our method then describes the swirling velocity of a core of an overall non-swirling vortex \citep[see also][]{Moffatt}.


 I would like to thank Jonathan Braithwaithe,  Martin Kruczenski,  Daniel Phillips, Andreas Reisenegger, Hendrik Spruit and Dmitri Uzdensky.

\bibliographystyle{apj}

\appendix

\section{Non-barotropic EoS}
\label{Non-barotropic}

In non-barotropic fluid
Eq. (\ref{MHD1}) becomes 
 \be
\nabla \times  {\J \times \B  \over \rho}= {\nabla p \times \nabla \rho \over\rho^2}
\label{MHD2}
\ee
Axial symmetry and the fluid approximation still  imply $I=I(P)$. 
 The force balance equation (\ref{MHD2}) 
 gives
 \be
\nabla P \times \nabla {\Delta ^\ast P +  { 4 I I'  } \over \varpi^2 \rho} = { \nabla p \times  \nabla \rho \over \rho^2}
  \label{GSnonbar}
  \ee
writing $p=p(s, \rho)$, where $s$ is entropy and assuming 
\be
 {\Delta ^\ast P +  { 4 I I'  } \over \varpi^2 \rho}= F(s, P), 
 \ee
 we find
 \be
 \nabla P \times \nabla F =  {\partial F \over \partial s} \nabla P \times \nabla s =   \left(  {\partial p \over \partial s} \right)_\rho { \nabla s \times  \nabla \rho \over \rho^2}
 \ee
 If we assume that distribution of density is spherically symmetric, $\rho =\rho(r)$, we find
 \be
 F=F(P) - \int  \left(  {\partial p \over \partial s} \right) _\rho {\partial_r \rho \over \partial_r P} {ds \over \rho^2}
 \label{FF}
 \ee
 This form of  $F$ should be used in the Grad-Shafranov equation (\ref{GS}), which makes it an integro-differential equation for $P$ \citep[\cf][Eq. 20]{VillataFerrari}.
Thus, in case of non-barotropic fluid the poloidal current is still function of $ P $ only, $ I=I(P)$, plus there will be an extra term in the Grad-Shafranov equation, Eq. (\ref{FF}). Thus, one still encounters  the same problem, that on the boundary of the enclosed toroid with toroidal \Bf\ both flux function and its derivative should match those in the bulk.

\section{Explicit example of expansion of $P$  near the point $\varpi=1$, $z=0$}
\label{expansion}

Let us illustrate the procedure of finding the flux function inside the toroid  starting from a  particular point on the separatrix, $\varpi=1$, $z=0$. In this case the normal coincides with the radial direction. We know the value of $\left.P_0\right|_\partial=1/2$ and its derivative $\partial_n \left.P_0\right|_\partial= -1/2$ at this point. In addition the flux function is zero on the separatrix, $G(1/2)=0$, while the pressure function  is unity, $F(1/2)=1$.
 From (\ref{GS}) we find  $\partial_\varpi^2 P$ :
 \be
 \partial_\varpi^2 P= -  \partial_z^2 P+ { \partial_\varpi P \over \varpi} - {15 \over 2} \varpi^2 F(P) +G(P)
 \label{M}
 \ee
On the boundary it is equal to
\be 
\left.  \partial_\varpi^2 P \right|_\partial = 
\left.
{1 \over \varpi ^2} - {15\over 2} \varpi ^2 F(1/2) - G(1/2)\right|_\partial= - {13\over 2}, 
\ee
Taking a derivative of Eq. (\ref{M}) with respect to $\varpi$ we find
\be 
\left.  \partial_\varpi^3 P \right|_\partial = -18 + {15\over 4} F'(1/2) + {1\over 2} G'(1/2)
\ee
Thus, expansion of the flux function near the point $\varpi=1$, $z=0$ reads
\be 
P={1\over 2} - {1\over 2}(\varpi-1) - {13\over 2}{(\varpi-1) ^2 \over 2} + \left( -18 + {15\over 4} F'(1/2) + {1\over 2} G'(1/2) \right) {(\varpi-1) ^3 \over 6}+...
\ee
It can be continued to arbitrary high order. Matching of the expansions of the flux functions from different points then gives the coefficients  $F^{(m)}(1/2)$ and
$G^{(m)}(1/2)$, thus determining simultaneously 
both the  flux function $P$ and functions $F$ and $G$.

\end{document}